\documentclass[journal=jacsat,manuscript=article]{achemso}
\usepackage[version=3]{mhchem} 
\usepackage[T1]{fontenc}       

\usepackage{amsmath}
\usepackage{amssymb}
\usepackage{graphicx}
\usepackage{afterpage}
\usepackage{amsfonts}
\usepackage{amssymb}
\usepackage{graphicx}
\usepackage{braket}
\usepackage{textcomp}

\title{Radiatively limited dephasing and exciton dynamics in MoSe$_2$ monolayers}

\author{Tomasz Jakubczyk}
\affiliation{Universit\'{e} Grenoble Alpes, Institut N\'{e}el,
F-38000 Grenoble, France} \alsoaffiliation{CNRS, Institut N\'{e}el,
"Nanophysique et semiconducteurs" group, F-38000 Grenoble, France}

\author{Valentin Delmonte}
\affiliation{Universit\'{e} Grenoble Alpes, Institut N\'{e}el,
F-38000 Grenoble, France} \alsoaffiliation{CNRS, Institut N\'{e}el,
"Nanophysique et semiconducteurs" group, F-38000 Grenoble, France}

\author{Maciej Koperski}
\affiliation{Laboratoire National des Champs Magn\'{e}tiques
Intenses, CNRS-UGA-UPS-INSA-EMFL, 25 Av. des Martyrs, 38042
Grenoble, France}\alsoaffiliation{Institute of Experimental Physics,
Faculty of Physics, University of Warsaw, ul. Pasteura 5, 02-093
Warsaw, Poland}

\author{Karol Nogajewski}
\affiliation{Laboratoire National des Champs Magn\'{e}tiques
Intenses, CNRS-UGA-UPS-INSA-EMFL, 25 Av. des Martyrs, 38042
Grenoble, France}

\author{Cl\'{e}ment Faugeras}
\affiliation{Laboratoire National des Champs Magn\'{e}tiques
Intenses, CNRS-UGA-UPS-INSA-EMFL, 25 Av. des Martyrs, 38042
Grenoble, France}

\author{Wolfgang Langbein}
\affiliation{Cardiff University School of Physics and Astronomy, The
Parade, Cardiff CF24 3AA, UK}

\author{Marek Potemski}
\affiliation{Laboratoire National des Champs Magn\'{e}tiques
Intenses, CNRS-UGA-UPS-INSA-EMFL, 25 Av. des Martyrs, 38042
Grenoble, France}

\author{Jacek Kasprzak}
\affiliation{Universit\'{e} Grenoble Alpes, Institut N\'{e}el,
F-38000 Grenoble, France} \alsoaffiliation{CNRS, Institut N\'{e}el,
"Nanophysique et semiconducteurs" group, F-38000 Grenoble, France}
\email{jacek.kasprzak@neel.cnrs.fr}
\begin{document}

\begin{abstract}
By implementing four-wave mixing (FWM) micro-spectroscopy we measure
coherence and population dynamics of the exciton transitions in
monolayers of MoSe$_2$. We reveal their dephasing times T$_2$ and
radiative lifetime T$_1$ in a sub-picosecond (ps) range, approaching
T$_2$=2T$_1$, and thus indicating radiatively limited dephasing at a
temperature of 6\,K. We elucidate the dephasing mechanisms by
varying the temperature and by probing various locations on the
flake exhibiting a different local disorder. At a nanosecond range,
we observe the residual FWM produced by the incoherent excitons,
which initially disperse towards the dark states, but then relax
back to the optically active states within the light cone. By
introducing polarization-resolved excitation, we infer inter-valley
exciton dynamics, showing an initial polarization degree of around
30\%, constant during the initial sub-picosecond decay, followed by
the depolarization on a picosecond timescale. The FWM hyperspectral
imaging reveals the doped and undoped areas of the sample, allowing
to investigate the neutral exciton, the charged one or both
transitions at the same time. In the latter, we observe the
exciton-trion beating in the coherence evolution indicating their
coherent coupling.
\end{abstract}

\newcommand{\Ea}{\ensuremath{{\cal E}_1}}
\newcommand{\Eb}{\ensuremath{{\cal E}_2}}
\newcommand{\Ec}{\ensuremath{{\cal E}_3}}
\newcommand{\Ei}{\ensuremath{{\cal E}_i}}
\newcommand{\Ed}{\ensuremath{{\cal E}_{1,\,2,\,3}}}
\newcommand{\Eg}{\ensuremath{{\cal E}_{1,\,2}}}
\newcommand{\Er}{\ensuremath{{\cal E}_{\rm r}}}

\paragraph{\textbf{Keywords:}}{excitons, coherent
nonlinear spectroscopy, transition-metal dichalcogenides, MoSe$_2$,
four-wave mixing, coherent nonlinear spectroscopy}

\paragraph{\textbf{Introduction}}
The identification of atomically-thin solids\,\cite{NovoselovSci04}
resulted in the development of the intriguing physics of graphene,
followed by the emerging technological
applications\,\cite{FioriNatNano14, KoppensNatNano14}. Also, it
stimulated a rapid progress in fundamental studies of thin films
extracted from other than graphite layered materials, such as,
semiconducting transition metal dichalcogenides (S-TMDs). The
bandgap of S-TMDs, converts from indirect to direct, when reducing
the material thickness to a single monolayer\,\cite{MakPRL10},
enabling exceptionally strong excitonic transitions. Owing to the
breakdown of out of plane translational symmetry for two dimensional
systems, the coupling of excitons with light is
boosted\,\cite{DeveaudPRL91}, resulting in their short radiative
lifetime and thus increased oscillator strength $\mu$. Excitons in
S-TMD monolayers display large binding energies E$_{\rm B}$ of
several hundreds of meV\,\cite{YeNature14, ChernikovPRL14,
WangPRL15, OlsenPRL16} - an increase by one to two orders of
magnitude with respect to a typical semiconductor quantum
well\,\cite{LeavittPRB90, AndreaniBook94}. The observed excitonic
absorption in S-TMDs reaches the values as high as
10\,\%\cite{AroraNanoscale15}, illustrating an exceptionally strong
$\mu$ in these systems and implying a radiative lifetime in the
sub-picosecond (sub-ps) range\,\cite{PalummoNL15}, as recently
observed in monolayers of
WSe$_2$\,\cite{PoellmannNatMat15,MoodyNatComm15} and other emerging
two-dimensional systems, namely nano-platelets\,\cite{NaeemPRB15}.
The robust coupling with light is attractive in prospective
applications, especially for
photodetectors\,\cite{KoppensNatNano14}. It is also appealing in
optical fundamental studies, for instance in
polaritonics\,\cite{LiuNatPhot14} and nonlinear
spectroscopy\,\cite{MoodyNatComm15, MalardPRB13, Hao2015}.

The large $\mu$ in S-TMDs gives rise to a giant nonlinear optical
response, which is investigated in this work via three-pulse
four-wave mixing (FWM) micro-spectroscopy. This approach offers a
direct access to the exciton coherence and population dynamics with
a time resolution limited only by the duration of the laser pulses,
which resonantly excite targeted optical transitions. Indeed,
time-resolved non-resonant photoluminescence
measurements\,\cite{LagardePRL14}, offer insufficient
time-resolution to infer sub-ps evolution and involve complex
relaxation pathways. Instead, resonant
experiments\,\cite{MoodyNatComm15, Hao2015}, also investigating
internal transitions\,\cite{PoellmannNatMat15} occurring in the THz
domain, have recently revealed ultrafast radiative recombination of
the exciton ground state in WSe$_2$.

In this work, we show that the optical dephasing time T$_2$ of the
neutral exciton transition (EX) in a monolayer of MoSe$_2$ is
intrinsically limited by the EX ultrafast radiative recombination
T$_1$. In contrast to previous works inferring optical coherence in
S-TMDs\,\cite{MoodyNatComm15, Hao2015}, we take advantage of the
micro-spectroscopy approach, employing the laser beams focused down
to the diffraction limited size of 0.7\,$\mu$m (full width at
half-maximum, FWHM). Such implementation is used to perform the FWM
hyperspectral imaging\,\cite{FrasNatPhot16}, which discriminates the
exciton charge state across the sample and helps to reveal the
striking features of coherent coupling between the neutral and
charged excitons. We demonstrate that the EX linear response
inferred via micro-reflectance, is affected by the inhomogeneous
broadening $\xi$ also on a sub\,-$\mu$m scale, evidenced by the
photon echo formation in the FWM transients\,\cite{LangbeinPRL05}.
The analysis of the FWM delay dependence yields T$_2$$\simeq$2T$_1$,
occurring on a sub-picosecond timescale. A spatially-resolved study
shows that T$_2$ depends on the local disorder, generating spatially
varying localization potentials for excitons and thus varying $\xi$,
directly influencing T$_1$. From the temperature dependent
homogenous linewidth, retrieved from the measured coherence
dynamics, we determine the dephasing due to phonon interaction,
which is described by a linear part and a thermally activated part,
as is quantum wells. We also monitor the density dynamics of
excitons, which is governed by the interplay between bright states
within the light cone and various available dark states.

\begin{figure}[!ht]
\includegraphics[width=0.85\textwidth]{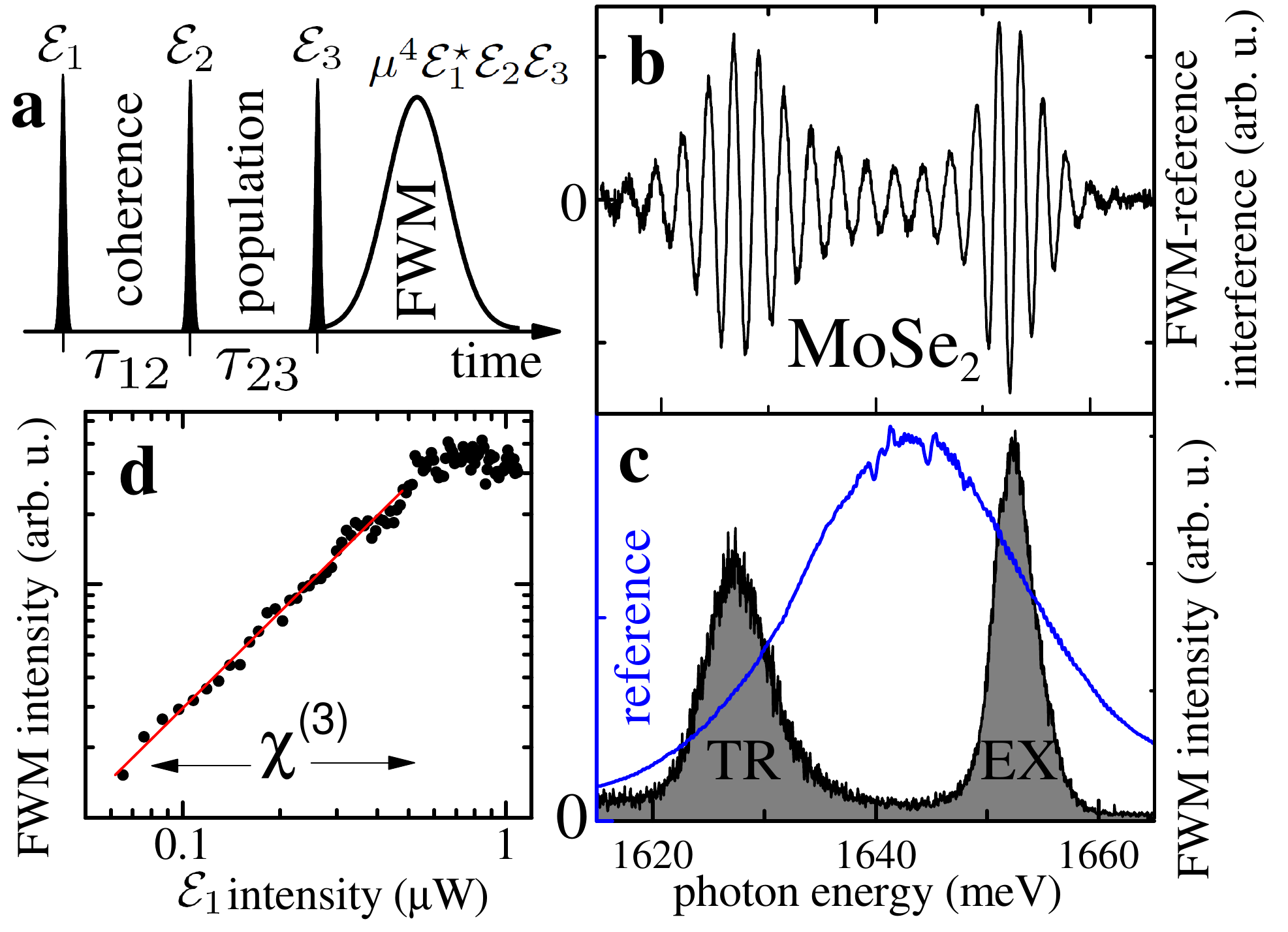}
\caption{{\bf Four-wave mixing spectroscopy of the MoSe$_2$
monolayer.} a)\,Pulse sequence employed in FWM experiments and
related observables. b)\,FWM spectral interferogram obtained on the
flake position displaying the exciton (EX) and trion (TR)
transitions. c)\,FWM intensity (black) retrieved from b) via
spectral interferometry. $\Er$ is shown with a blue line.
d)\,Spectrally integrated FWM intensity of EX as a function of $\Ea$
intensity, showing the driving range yielding the $\chi^{(3)}$
regime of the FWM.\label{fig:fig0}}
\end{figure}

\begin{figure}[!ht]
\includegraphics[width=0.71\textwidth]{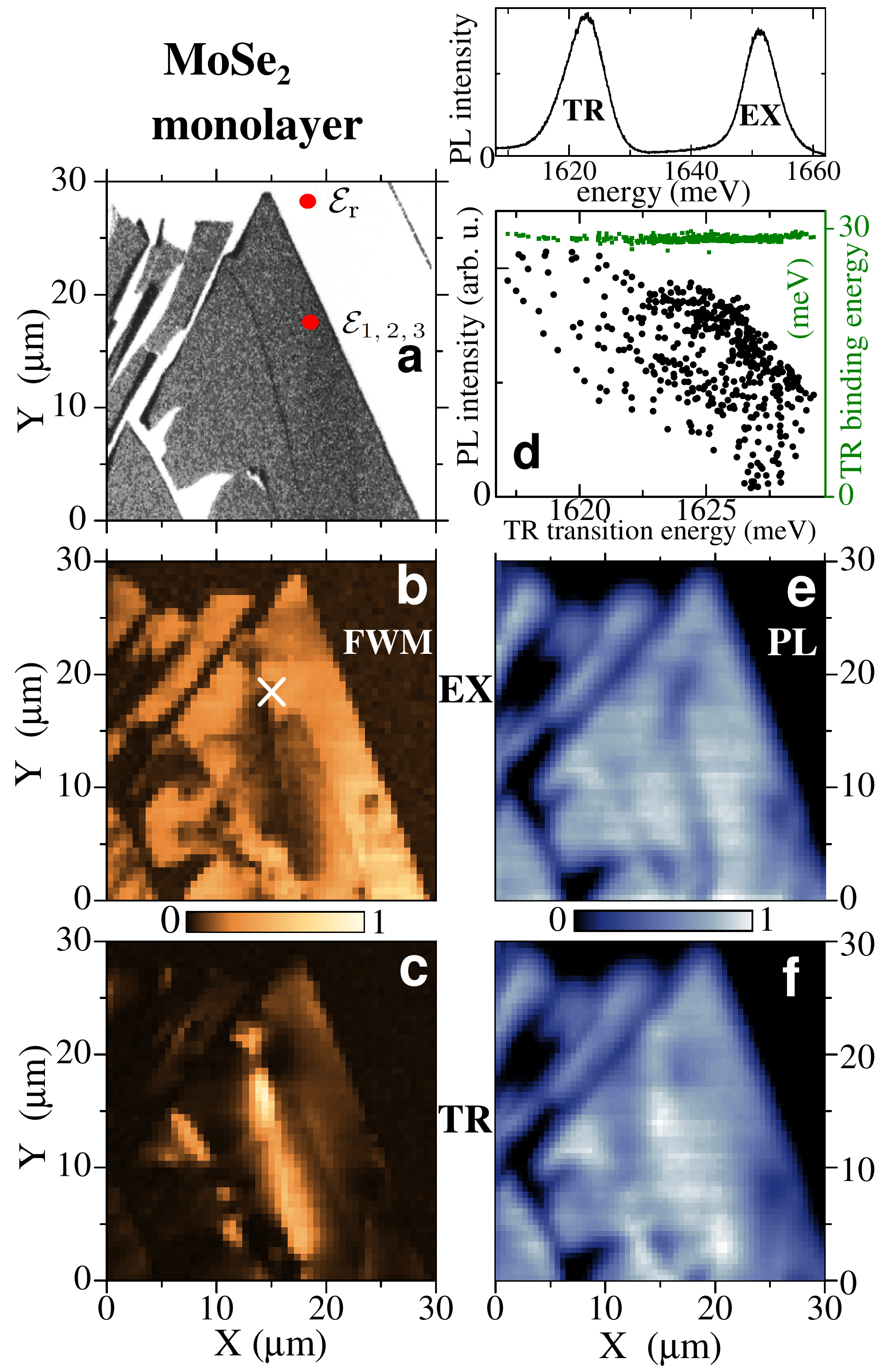}
\caption{\textbf{Hyperspectral mapping of the MoSe$_2$ monolayer.}
a)\,Optical contrast of the sample observed in reflectance.
b)-c)\,Spatial imaging of the FWM amplitude, spectrally averaged
over the exciton (b) response, centered at $\sim1650\,$meV and the
trion (c) at $\sim1625\,$meV. Heterodyning at the FWM frequency
$\Omega_3+\Omega_2-\Omega_1$ with $\tau_{12}=\tau_{23}=50\,$fs,
T$=6\,$K. d)\,Top: A typical PL spectrum, non-resonantly excited at
at $\sim1750\,$meV ($\sim710\,$nm) with $\simeq0.1\,\mu$W average
power arriving at the sample surface, displaying bright emission
from EX and TR, with integrated count rate of 350kHz for each
transition. Bottom: Correlation between the PL intensity of the TR
(black) and its binding energy (green), as a function of its
transition energy. e)-f)\,PL imaging of TR and EX, respectively.
Excitation conditions as in d). Linear color scale, as shown by
horizontal bars.\label{fig:fig_imagerie}}
\end{figure}

\paragraph{\textbf{Sample \& experiment}} An image of the studied MoSe$_2$ monolayer is shown in
Fig.\,\ref{fig:fig_imagerie}. It was exfoliated from bulk material
and transferred onto a Si/SiO$_2$ substrate, with a nominal SiO$_2$
thickness of 86\,nm. Relatively large monolayer flakes of up to
$(50\,\times\,50)\,\mu$m$^2$ size have been fabricated and placed in
an optical He-flow cryostat. To infer both the coherence and the
population dynamics of excitons, we retrieved their FWM by
implementing a three-beam configuration\,\cite{FrasNatPhot16} of the
heterodyne spectral interferometry\,\cite{LangbeinPRL05}. This
technique has been proven as an efficient detection scheme of
optical nonlinearities in a solid, until now only employed to
retrieve wave mixing signals generated by individual transitions in
semiconductor quantum dots\,\cite{LangbeinPRL05, KasprzakNPho11,
FrasNatPhot16}. FWM is an optical polarization created with short,
resonant driving pulses; $\Ea$, $\Eb$ and $\Ec$, as depicted in
Fig.\,\ref{fig:fig0}\,a. In the third-order $(\chi^{(3)})$ regime
its amplitude is proportional to $\mu^4\Ea^{\ast}\Eb\Ec$. Thus,
owing to a large $\mu$, a dramatic enhancement of FWM is expected in
MoSe$_2$ monolayers. Note that the $^{\ast}$ stands for the complex
conjugate, which is the origin of the FWM rephasing in
inhomogeneously broadened systems\,\cite{LangbeinPRL05}, generating
the photon echo (as also sketched in Fig.\,\ref{fig:fig0}\,a), which
acts as a probe of the microscopic dephasing. Micro-FWM spectroscopy
requires the co-linear arrangement of the driving fields $\Ed$,
which is enabled by phase-selecting the signal through optical
heterodyning\,\cite{LangbeinPRL05}. By employing acousto-optic
modulation, $\Ed$ are frequency up-shifted by radio-frequencies,
introducing controlled phase-drifts in their respective pulse trains
generated by a Ti:Sapphire femto-second laser. As we intend to
measure sub-ps dynamics, pulseshaping\,\cite{FrasNatPhot16} has been
applied to correct the temporal chirp. After acquiring the delays
$\tau_{12}$ and $\tau_{23}$, introduced by a pair of mechanical
delay lines, $\Ed$ are recombined into a common spatial mode and are
focused on the sample with the microscope objective. The reference
beam $\Er$, used in the heterodyne mixing and interferometric
detection, is also focused at the sample, yet is displaced with
respect to $\Ed$, as depicted in Fig.\,\ref{fig:fig_imagerie}\,a.
The time-ordering of the pulses is presented on
Fig.\,\ref{fig:fig0}\,a: measuring time-integrated FWM detected at
the $\Omega_3+\Omega_2-\Omega_1$ heterodyne frequency, as a function
of $\tau_{12}$ ($\tau_{23}$), yields the coherence (population)
dynamics of an optical transition. The FWM signal is measured in
reflectance, attaining a shot-noise detection limit and rejecting
the resonant driving fields $\Ed$ with $10^{6}$ ($10^{12}$)
selectivity in field (intensity). The interference between the
heterodyned signal and $\Er$ is spectrally resolved with an imaging
spectrometer. Further details regarding the current experimental
implementation are given in Ref.\,[20].

In Fig.\,\ref{fig:fig0}\,b we present a typical spectral
interference between $\Er$ and the FWM at a temperature T=5\,K for
$\tau_{12}=0$. The FWM intensity is retrieved by spectral
interferometry and shown in Fig.\,\ref{fig:fig0}\,c. The signal
consists of two transitions identified as the ground
state\,-\,exciton (EX) and the single electron\,-\,trion transitions
(TR)\,\cite{WangAPL15}. In Fig.\,\ref{fig:fig0}\,d we present the
FWM intensity of EX as a function of $\Ea$ intensity demonstrating a
linear dependence in the $\chi^{(3)}$ regime, as expected. Note that
the FWM can be driven with a $\Ea$ intensity as low as a few tens of
nW, corresponding to a few hundreds of photons per pulse $\Ea$ and
generating a low carrier density, less than $10^{9}/$\,cm$^{2}$.
Such density is far below the saturation
density\,\cite{JonesNatNano13} estimated at around
$10^{13}/$\,cm$^{2}$. In our study, we therefore consider the
creation of excitons with K$\sim0$ center of mass momentum, which
appear at the, K+ and K-, points of the Brillouin zone of the S-TMD
crystal. Those excitons can either decay radiatively or disperse out
of the light cone (K$>n\omega/c$) via phonon scattering. The latter
process occupies dark exciton states, which relax back to
K$<n\omega/c$ on a pico-second time scale and eventually recombine.
Further relaxation pathways are scattering electrons and holes
between the K-points.

\paragraph{\textbf{Hyperspectral imaging}}
The micro-spectroscopy approach enables to perform FWM hyperspectral
imaging\,\cite{KasprzakNPho11, FrasNatPhot16}, as shown in
Fig.\,\ref{fig:fig_imagerie}. It allows to identify regions of the
flake dominated by the FWM of EX or TR (see
Fig.\,\ref{fig:fig_imagerie}\,b and c). Clearly, the two images are
complementary, which permits to distinguish regions of different
resident carrier concentration. In Fig.\,\ref{fig:fig_imagerie}\,b
we note that the FWM amplitude remains virtually constant over the
areas of more than $(5\,\times 5)\,\mu$m$^2$, indicating weak
disorder and thus enabling an extremely fast radiative decay of
excitons. In the following experiments, the performed imaging
allowed us to selectively address EX or TR, or to drive
simultaneously both transitions. Performing micro-spectroscopy also
permits to locally address sub\,-$\mu$m regions of smaller spectral
inhomogeneous broadening with respect to the total area of the
flake. To exemplify this, we have performed a statistical analysis
of the micro-photoluminescence (PL) hyperspectral imaging. Similarly
as in Fig.\,\ref{fig:fig0}\,c, confocally detected PL spectra yield
EX and TR transitions, as displayed at the top of
Fig.\,\ref{fig:fig_imagerie}\,d. We observe particularly bright
emission, with spectrally integrated count rate of typically
350\,kHz from each transition. For every spatial position, we have
determined the integrated intensity for both transitions and their
center energies. For the latter, we observe the spread over
$\xi\simeq10\,$meV, as displayed at the bottom of
Fig.\,\ref{fig:fig_imagerie}\,d. Interestingly, despite this large,
macroscopic $\xi$ - which we attribute to the strain distribution
across the flake - the trion binding energy $\Delta$ remains well
defined, $\Delta=(28.8\,\pm0.3)\,$meV. We note that the lower the TR
transition energy, the higher its intensity is measured, which is
attributed to the the distribution of residual electrons in the
sample. Interestingly, such correlation is not observed for EX,
i.e., the EX intensity is not sensitive to the apparent, in our
sample changes in the charge density. In
Fig.\,\ref{fig:fig_imagerie}\,e and f we present the PL imaging
spectrally integrated over EX and TR transition, respectively. By
comparing it with Fig.\,\ref{fig:fig_imagerie}\,b and c, we point
out two advantages of the FWM imaging with respect to the PL one.
Firstly, owing to the third-power scaling of the FWM intensity with
the excitation power, combined with heterodyning with $\Er$, the
spatial resolution in the FWM imaging is
enhanced\,\cite{LangbeinRNC10} up to $0.3\lambda/{\rm NA} \approx
360$\,nm, surmounting the standard diffraction limit by a factor of
2. Secondly, the FWM yields a significantly improved imaging
selectivity of EX and TR across the flake. In fact, different
properties are inferred in both experiments. The FWM directly probes
$\mu$ of the resonantly generated excitons at the K-points of the
valleys. Instead, the non-resonant PL reflects more complex carrier
relaxation along the valleys towards their K-points, prior to the
exciton formation, followed by their radiative recombination.

\begin{figure}[!ht]
\includegraphics[width=0.85\textwidth]{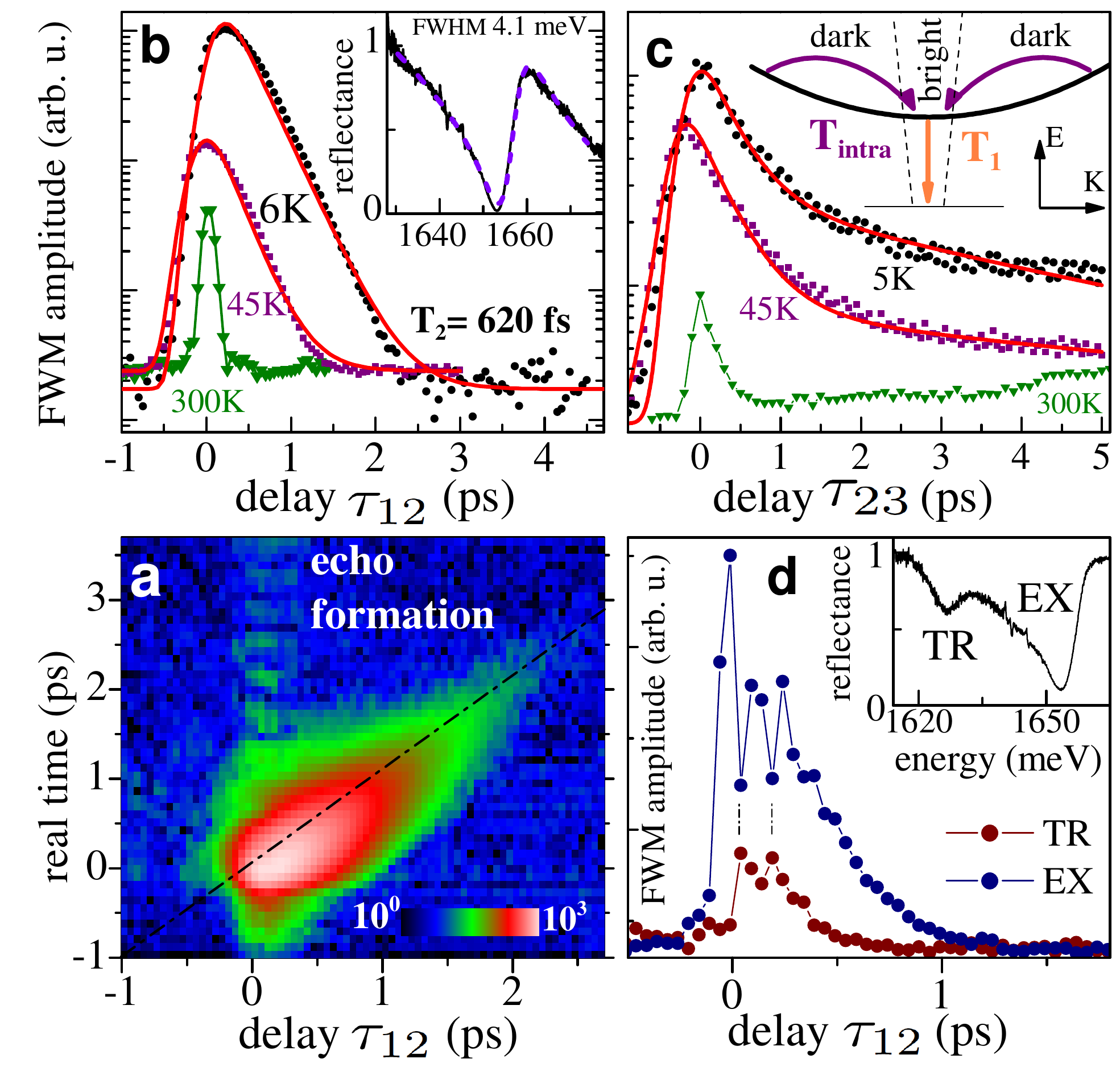}
\caption{\textbf{Exciton dynamics in the MoSe$_2$ monolayer inferred
with FWM microscopy}. a)\,Time-resolved FWM amplitude as a function
of $\tau_{12}$, showing formation of a photon echo: a Gaussian
signal with a maximum for $t=\tau_{12}$. Logarithmic color scale
given by the color bar. b)\,FWM amplitude as a function of
$\tau_{12}$ retrieved from the EX at T$=(6,\,45,\,300)\,$K given by
(black circles, purple squares and green triangles), respectively.
The simulations yielding T$_2\simeq(620,\,520)\,$fs are given by red
traces. Inset:\,reflectance measured at T=6\,K yielding $4.1\,$meV
FWHM. c)\,Cartoon of the considered radiative recombination and
intra-valley relaxation processes is presented in the inset. The
initial dynamics of the EX population measured at (5,\,45,\,300)\,K,
color coding as in b). The results yield radiative lifetime T$_1$
and relaxation time $T_{\rm intra}$ (see main text). d)\,$\tau_{12}$
dependence of the FWM amplitude when simultaneously driving the EX
(blue) and TR (brown) transitions, revealing a beating with 140\,fs
period, and thus indicating EX-TR coherent coupling. The location of
the excitation is marked with a cross in
Fig.\,\ref{fig:fig_imagerie}\,b, while the corresponding
micro-reflectance spectrum is given in the
inset.\label{fig:fig2coherence}}
\end{figure}

\paragraph{\textbf{Coherence dynamics}}
The strength of the FWM spectroscopy in assessing the coherence in
solids lies in its capability to separate homogenous ($\gamma$) and
inhomogeneous ($\xi$) contributions of the transition's spectral
width. In particular, in a presence of $\xi$, the time-resolved FWM
amplitude exhibits a photon echo\,\cite{MoodyNatComm15, NaeemPRB15},
which decays as $\exp{(-2\tau_{12}/{\rm T}_2)}$. Hence, to
investigate the exciton coherence dynamics, we measured degenerate
FWM $(\Omega_2=\Omega_3)$, as a function of $\tau_{12}$.
Time-resolved FWM amplitude of the EX transition, displayed in
Fig.\,\ref{fig:fig2coherence}\,a, clearly demonstrates formation of
the photon echo. From its temporal width we estimate the local $\xi$
to be in the meV range (around 3\,meV for the case shown in
Fig.\,\ref{fig:fig2coherence}\,a). Fig.\,\ref{fig:fig2coherence}\,b
shows the time-integrated FWM as a function of $\tau_{12}$. The data
are modeled by a convolution of a Gaussian profile with an
exponential decay. The former exhibits the FWHM width of 0.16\,ps,
reflecting duration of $\Ed$ impinging the sample. Instead, from the
latter we retrieve at T=6\,K the dephasing time
T$_2=2\hbar/\gamma$=$(620\,\pm\,20)\,$fs, and thus
$\gamma\simeq2.1\,$meV (FWHM). For simplicity, the dynamics owing to
the echo formation process close to zero delay, has been here
disregarded, yet it explains a minor deviation between the
measurement and prediction. The homogenous broadening $\gamma$ is
around twice smaller than the transition linewidth directly measured
via micro-reflectivity, as shown in the inset of
Fig.\,\ref{fig:fig2coherence}\,b and also in FWM (see
Fig.\,\ref{fig:fig0}\,c). We thus conclude that the line-shape
remains affected by the inhomogeneous broadening $\xi$, even though
a sub-$\mu$m area is probed. With increasing temperature, T$_2$ is
expected to decrease\,\cite{MoodyNatComm15}, owing to phonon
scattering. This is highlighted in Fig.\,\ref{fig:fig2coherence}\,b.
At T=45\,K the dephasing accelerates and we measure
T$_2=(520\,\pm\,40)\,$fs. At ambient temperature the dephasing
occurs at a timescale faster than 100\,fs and is not resolved by our
setup, although a pronounced FWM is still measured. We note that for
the TR transition at T=6\,K we find T$_2=(460\,\pm\,30)\,$fs and
similar inhomogeneous broadening as for EX, also generating a photon
echo as in Fig.\,\ref{fig:fig2coherence}\,c (not shown). This
shorter dephasing of the charged exciton, is tentatively interpreted
in terms of the final state damping, due to the energy distribution
of final state energies of the leftover electron. In
Fig.\,\ref{fig:fig2coherence}\,c we present the coherence dynamics
measured at the boundary of the doped and undoped regions of the
flake, marked with a cross in Fig.\,\ref{fig:fig_imagerie}\,d, such
that the FWM of both EX and TR is driven in tandem. We observe a
beating, withstanding on both transitions during initial positive
delays $\tau_{12}$, up to around 700\,fs when the TR coherence
virtually vanishes. The beating period of $\zeta=140\,$fs (marked
with a pair of vertical lines), well corresponds to the EX\,-\,TR
binding energy $\Delta=2\pi\hbar/\zeta\simeq29\,$meV and thus
indicates their coherent coupling\,\cite{SinghPRL14}. To support
this statement - and in particular, to distinguish between
polarization interference and coherent
coupling\,\cite{KasprzakNPho11} - the data have been
Fourier-transformed along the delay $\tau_{12}$ to yield the
two-dimensional FWM spectrum, indeed revealing off-diagonal coupling
terms between EX and TR (not shown).

\begin{figure}[!ht]
\includegraphics[width=0.9\textwidth]{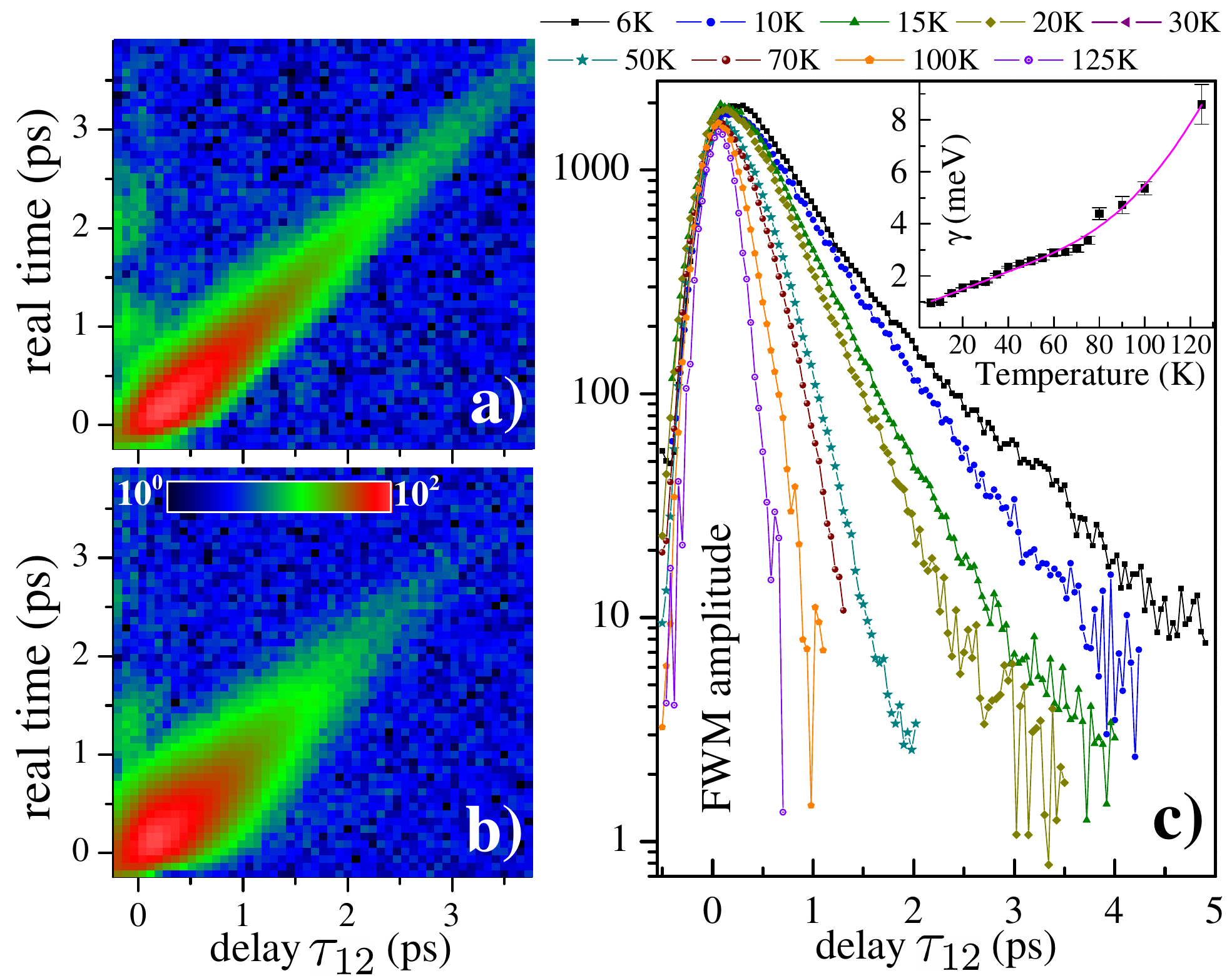}
\caption{\textbf{Impact of a local disorder and temperature on the
exciton dephasing in the MoSe$_{2}$ monolayer}. a)\,Time-resolved
FWM amplitude, measured on the location displaying a larger $\xi$,
showing a temporarily narrowed photon echo. The observed increased
dephasing time with respect to Fig.\,\ref{fig:fig2coherence}
attributed to a localization induced increase of the radiative
lifetime T$_1$. b)\, as a) but measured on the area showing a
broader echo and thus smaller $\xi$. A weaker localization yields a
shorter T$_1$=T$_2$/2 than in a). Logarithmic color scale over two
orders of magnitude. c)\,FWM amplitude dynamics as a function of
temperature measured on the area as in a). Above T=125\,K the
dephasing is faster than the temporal resolution of the experiment.
The temperature dependent homogenous broadening $\gamma=2\hbar/{\rm
T} _{2}$ (inset) indicates the phonon-induced dephasing
mechanism.\label{fig:figmechanisms}}
\end{figure}

To gain a deeper understanding  of the exciton ultrafast dynamics,
we have combined the enhanced spatial and temporal resolution of our
experiment and we have performed spatially-resolved dephasing study
at T=6\,K. Within an area of 8\,$\mu$m by 8\,$\mu$m, displaying
uniquely the EX transition, we have scanned the coherence dynamics
with a spatial step of $0.66\,\mu$m. The analysis of the obtained
statistics of dephasing (169 traces), reveals variations of T$_2$
within the probed area from around 0.5\,ps up to 1.5\,ps.
Interestingly, the locations on the flake yielding the shortest
T$_2$ also display the broadest photon echo, and thus the smallest
$\xi$. Conversely, the longest T$_2$ is measured on the areas
characterized by a larger $\xi$, and thus showing the narrowest
photon echo, here limited by the temporal duration of the laser
pulses. An example of such two representative cases is displayed in
Fig.\,\ref{fig:figmechanisms}\,a and b. The spatially-resolved
dephasing experiment indicates that T$_2$ is governed by a local
disorder, realizing various localization potentials. The resulting
spatially-dependent coherence volume of EX has a direct impact on
its radiative lifetime T$_1$: the fastest recombination is expected
at the areas of the smallest $\xi$, as indeed measured. Full spatial
correlations between T$_1$, T$_2$ and $\xi$ will be reported in a
forthcoming publication. In Fig.\,\ref{fig:figmechanisms}\,c we
present the coherence dynamics on the chosen area exhibiting an
increased T$_2$, measured from 6\,K to 150\,K. The data clearly show
a gradual decrease of T$_2$ with temperature from 1.4\,ps to 0.2\,ps
respectively. The retrieved $\gamma=2\hbar/{\rm T}_2$ is plotted in
the inset.  It can be modeled with a linear dependence and an
additional bosonic term\,\cite{HorzumPRB13}: $\gamma({\rm
T})=\gamma_0+a{\rm T}+b/(\exp(E_1/k_{B}{\rm T})-1)$. The linear term
($\gamma_0=(0.78\,\pm\,0.11)\,$meV, $a=(0.03\,\pm\,0.003)\,$meV/K)
is due to low energy acoustic phonons. The latter term, with the
energy $(E_1=43\,\pm\,4)\,$meV and $b=(187\,\pm\,75)\,$meV, could be
attributed to thermal activation of higher energy optical
phonons\,\cite{HorzumPRB13}.

\paragraph{\textbf{Population dynamics}} In the following, the FWM is employed to infer the EX population
dynamics after their resonant and selective excitation in a given
valley ($\Ea$ and $\Eb$, denoted as $\Eg$, are co-circularly
polarized). They arrive at the flake with virtually no delay,
$\tau_{12}=(50\pm10)\,$fs, generating exciton population. Owing to
the large $\mu$, the excitons exhibit fast radiative decay, yet also
are subject to scattering on lattice defects and phonons. The matrix
element for the Coulomb induced, parametric exciton-exciton
scattering is proportional to the square of an exciton Bohr
radius\,\cite{PorrasPRB02}. We thus propose that the Coulomb
scattering is reduced for these spatially compact excitons. In
particular, this channel is negligible for the small exciton
densities employed here. Nevertheless, excitons still experience
scattering\,\cite{LangbeinPRL02a} for instance induced by defects,
disorder and phonons, which can efficiently redistribute them out of
the light cone (marked with dashed lines in the inset of
Fig.\,\ref{fig:fig2coherence}\,c), toward K$>n\omega/c$. This type
of scattering is particularly efficient in S-TMDs: due to heavy
masses (and thus flat bands at the bottom of the branches) and large
$\gamma$, the excitons' center of mass scatter out of the light cone
at practically no cost in energy. As a result, one part of the
created population instantly decays with its radiative lifetime
T$_1$, while the remaining part spreads along the dispersion branch
populating dark states (i.e. outside the light cone) or is scattered
into one of the counter-polarized valleys involving spin-flip
processes. Also a part of these excited excitons eventually relax
toward the bottom of the valley with a characteristic time T$_{\rm
intra}$ and recombine radiatively, as depicted in
Fig.\,\ref{fig:fig2coherence}\,c. In the dynamics probed with FWM,
we disregard the influence of strongly localized, quantum-dot like
states, due to their small density (not detected on the region shown
in Fig.\,\ref{fig:fig_imagerie}). For simplicity, we also disregard
non-radiative Auger processes\,\cite{PoellmannNatMat15}, as we
operate at low exciton densities.

The FWM, triggered by $\Ec$ from the density grating generated by
$\Ea$ and $\Eb$, therefore probes all the above mentioned processes
via $\tau_{23}$ dependence. This is shown in
Fig.\,\ref{fig:fig2coherence}\,c, for the EX (obtained the same
spatial position as in Fig.\,\ref{fig:fig2coherence}\,b) for driving
upon co-circular polarization of $\Ed$. The data are modeled with a
double exponential decay convoluted with the Gaussian laser pulse,
showing temporal width of 0.16\,ps (FWHM). From the initial decay we
retrieve the exciton lifetime of T$_1=(390\,\pm\,20)\,$fs, while the
FWM for further delays $\tau_{23}$ yields the intra-valley
relaxation time of T$_{\rm intra}=(4.3\,\pm\,0.6)\,$ps. At T=45\,K
we measure instead (T$_1$,\,T$_{\rm
intra}$)=$(0.42\,\pm\,0.01,\,6.84\pm\,0.38)\,$ps. An increase of
T$_{\rm intra}$ with temperature is attributed to Boltzmann
distribution of excitons, allowing for reaching higher energies and
K-vectors out of the light cone, and also to access different dark
states offered by the complex structure of the valley-excitons. A
physical picture arising from the FWM experiment at low temperature
is that the initial exciton decay with the time T$_1$, covering an
order of magnitude in amplitude (see
Fig.\,\ref{fig:fig2coherence}\,c), is due to the radiative
recombination, while the non-radiative processes are of minor
impact. Importantly, comparing the T$_1$ with T$_2$ times, we
conclude that the dephasing is principally due to the radiative
decay, nearing to the radiative limit - the data can also be well
modeled by fixing T$_2$=2T$_1$ and using T$_{\rm intra}$ as the only
fitting parameter. Observation of the radiatively limited dephasing
is a prerequisite for implementing more advanced optical coherent
control schemes in S-TMDs.

We point out that the dynamics of the secondary, incoherent excitons
- here probed by $\tau_{23}$ dependence of FWM - is particularly
complex and might be influenced by a set of unconventional features
present in S-TMDs. One should bear in mind coupling with and
scattering toward counter-polarized K-valleys and spin-split bands,
relevant in formation of exciton complexes. In fact,
triple-degeneracy of the K-valleys, enables various configurations
for bright and dark excitons states, as previously considered for
other systems hosting multi-excitons\,\cite{AnNL07,MasiaPRB11}. We
also point out exotic dispersion relations for the center-of-mass
momentum\,\cite{QiuPRL15}. Thus, we refrain from firm interpretation
and first-principle modeling of the secondary exciton dynamics,
displayed in Fig.\,\ref{fig:polarres} on a 100\,ps time scale.

\paragraph{\textbf{Inter-valley dynamics}} Below, we present the FWM results obtained upon
polarization-resolved driving, employed to investigate exciton
scattering between counter-polarized valleys, testing the robustness
of the pseudo-spin degree of freedom. The reciprocal space of TMDs
monolayers displays nonequivalent bands with the extrema at
$K$-points\,\cite{XiaoPRL2012,JonesNatNano2013,XuNatPhys14}, labeled
as $K_+$ and $K_-$. The dipole-allowed transitions in these two
valleys can be selectively addressed by circularly polarized light,
$\sigma^+$ and $\sigma^-$, respectively. The light helicity is
therefore considered as an asset to drive, manipulate and read the
state of the valley subspaces. The valley
polarization\,\cite{JonesNatNano2013}, although protected by the
strong spin-orbit splitting in the valence band, decays mainly due
to the to electron-hole exchange
interaction\,\cite{YuPRB14,GlazovPRB14}. Since the stability of this
degree of freedom is required in prospective applications of TMDs,
intense efforts are currently devoted to study its dynamics and to
reveal the mechanisms that govern it\,\cite{Smolenski16}, in
particular in a presence of spin-forbidden
transitions\,\cite{AroraNanoscale15, ZhangPRL15, WangNatComm2015,
Echeverry2016, WithersNanoLett15}.

To infer the exciton inter-valley dynamics, we implemented
polarization-resolved excitation of the FWM signal. The polarization
state of the beams is adjusted by a set of $\lambda/2$ and
$\lambda/4$ plates, by monitoring spectral interferences of $\Ed$
with $\Er$. Specifically, $\Ec$ and $\Er$ are set as co-circular and
thus yield a maximal contrast of the heterodyne spectral fringes.
Instead, the polarization state of $\Eg$ is rotated until the
interference contrast is minimized, yielding opposite circular
polarizations for $\Eg$ (creating exciton population in the $K_+$)
and $\Ec$ (converting the population, redistributed from $K_+$ into
$K_-$, toward the FWM). To probe the exciton dynamics in the same
valley (as in Fig.\,3\,c of the manuscript), we set the co-circular
polarization of $\Eg$ and $\Ec$, by maximizing the spectral
interference of $\Eg$ and $\Er$. We estimate the polarization
cross-talk between $\Eg$ and $\Ec$ around $6\times10^{-4}$. Thus,
polarization selective driving permits to probe the dynamics of the
exciton transfer towards a counter-polarized valley, as depicted in
Fig.\,\ref{fig:polarres}, and thus to test the robustness of the
valley pseudo-spin\,\cite{XuNatPhys14}.

The measurement is presented in Fig.\,\ref{fig:polarres} (bottom).
As in Fig.\,3\,c, we observe an initial fast decay of the
radiatively recombining excitons within the first ps and subsequent
recombination of the secondary excitons. The early dynamics for both
polarization configuration is not the same during initial several ps
of $\tau_{23}$. Instead, for longer delays, $\tau_{23}>10\,$ps, the
measured FWM (generated by the secondary excitons) is the same for
both driving configurations. We calculate the FWM circular
polarization degree, which we define as
$\rho(\tau_{23})=\frac{R(\tau_{23},\sigma+)-R(\tau_{23},\sigma-)}{R(\tau_{23},\sigma+)+R(\tau_{23},\sigma-)}\times100\%$,
where $R$ denotes the spectral FWM amplitude. The resulting
$\rho(\tau_{23})$ is plotted at the top of Fig.\,\ref{fig:polarres}.
We thus do observe a significant pseudo-spin polarisation around
zero delay, $\rho(\tau_{23}\simeq0)\simeq30\%$, which however
vanishes extremely rapidly, within around 5\,ps, such that
$\rho(\tau_{23})=0$ for $\tau_{23}>10\,$ps. This result brings a new
input (demonstration of a possibility to create valley-polarized
excitons, but their rapid depolarization) towards a firm
understanding of an intriguingly weak efficiency of optical pumping
in MoSe$_2$ monolayers (in contrast to other
S-TMDs)\,\cite{DeryPRB15}. The polarization degree, relevant for
valleytronics applications, could be stabilized by using magnetic
fields\,\cite{Smolenski16} or by fabricating more involved
hetero-structures\,\cite{RiveraNatComm15, Rivera16} based on S-TMDs.
When analyzing Fig.\,\ref{fig:polarres} it is worth to note a large
fraction of the FWM response within initial several ps for both
polarization configurations, as compared to the subsequent dynamics.
This shows that the exciton density is mainly removed radiatively
within the sub-ps radiative lifetime T$_1$, as discussed before.

\paragraph{\textbf{Conclusions}} By performing FWM spectroscopy, we have
demonstrated a giant, optical, coherent, nonlinear response of
exciton transitions in MoSe$_2$ monolayers. We find an increase in
the FWM amplitude by two orders of magnitude, as compared to a high
quality CdTe semiconductor quantum well (not shown), emitting at the
similar wavelength and driven with a comparable excitation power.
Taking advantage of the microscopy approach, we have performed the
hyperspectral imaging, which allowed us to accurately determine the
areas yielding FWM responses of either neutral, or charged excitons,
or both. Electrical gating of such structures could enable spectral
control of coherent responses from TMDs, providing insights into the
mutual influence of TR and EX onto their dynamics. Using two-beam
FWM micro-spectroscopy, we have measured the excitons' coherence
evolution, accessing dephasing times T$_2$, which turn out to be
intrinsically limited by the radiative lifetime. The inhomogeneous
broadening $\xi$ on a sub-$\mu$m range is reduced by an order of
magnitude with respect to $\xi$ measured on the entire size of the
flake, although still gives rise to the pronounced photon echo. We
have also shown that T$_2$ spatially varies across the flake, as the
T$_1$ is governed by the local disorder, determining the exciton
localization. Prospective experiments, exploiting two-dimensional
FWM spectroscopy, will ascertain coupling mechanisms between exotic
valley-exciton species (excitons, trions, biexcitons and
multiexcitons) offered by S-TMDs. Our approach could be used to
retrieve coherent responses of the localized exciton states in
S-TMDs\,\cite{KoperskiNN2015}, which are expected to exhibit
ultra-long dephasing\,\cite{SchaibleyPRL15}. Finally, by exploiting
polarization-resolved, three-beam FWM we assessed the exciton
population dynamics, revealing the interplay between the sub-ps
radiative decay T$_1$ and the intra-valley relaxation T$_{\rm
intra}$ of the excitons scattered towards the dark states,
revisiting the light-matter coupling in two-dimensional
systems\,\cite{DeveaudPRL91}.

\emph{We acknowledge the support by the ERC Starting Grant PICSEN
contract no.\,306387 and the ERC Advanced Grant MOMB contract
no.\,320590. We thank G. Nogues and M. Richard for helpful remarks
on the manuscript. We also acknowledge the support from Nanofab
facility of the Institute N\'{e}el, CNRS UGA. Monolayers of MoSe$_2$
were obtained by means of polydimethylsiloxane-based exfoliation of
bulk crystals purchased from HQ Graphene.}

\providecommand{\latin}[1]{#1}
\providecommand*\mcitethebibliography{\thebibliography} \csname
@ifundefined\endcsname{endmcitethebibliography}
  {\let\endmcitethebibliography\endthebibliography}{}

\newpage
\textbf{Figures and captions:}

\begin{figure}[!ht]
\includegraphics[width=0.7\textwidth]{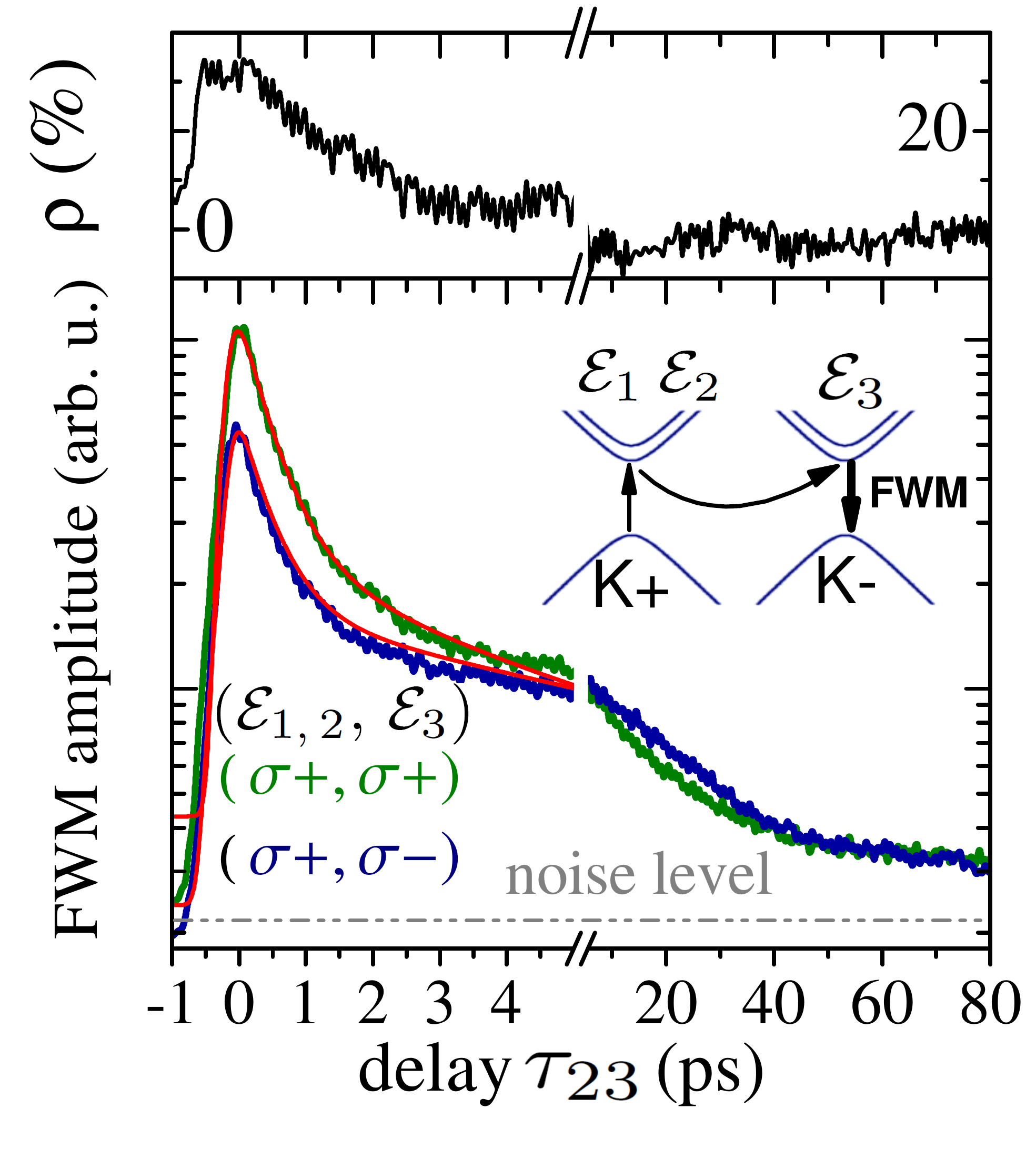}
\caption{\textbf{Inter-valley population dynamics in the MoSe$_2$
monolayers investigated with the polarization-resolved driving of
the FWM}. Bottom:\,FWM amplitude versus $\tau_{23}$ of the EX
transition obtained at T=$6\,$K upon co- (green circles) and
cross-circular (blue squares) setting of $(\Eg,\,\Ec)$. The
modeling, as in Fig.\,3\,c, is displayed as red lines. Equivalent
dynamics have been measured (not shown) for a counter-polarized
driving $(\sigma+\mapsto\sigma-)$. Inset:\,Rationale of the
inter-valley dynamics probed with FWM using polarization-resolved
driving: $\Ec$ induces the FWM signal in a given valley, which
originates from the population generated in the counter- or
co-polarized valley, by setting counter- or co-circular $\Eg$,
respectively. Top:\,$\tau_{23}$ delay dependence of the measured
circular polarization degree, $\rho(\tau_{23})$. We measure
$\rho(\tau_{23})$ of several tens of $\%$ for initial delays
$\tau_{23}$, followed by a total depolarization of the FWM,
generated by the secondary (see the main text) exciton population.
\label{fig:polarres}}
\end{figure}


\begin{mcitethebibliography}{44}
\providecommand*\natexlab[1]{#1}
\providecommand*\mciteSetBstSublistMode[1]{}
\providecommand*\mciteSetBstMaxWidthForm[2]{}
\providecommand*\mciteBstWouldAddEndPuncttrue
  {\def\EndOfBibitem{\unskip.}}
\providecommand*\mciteBstWouldAddEndPunctfalse
  {\let\EndOfBibitem\relax}
\providecommand*\mciteSetBstMidEndSepPunct[3]{}
\providecommand*\mciteSetBstSublistLabelBeginEnd[3]{}
\providecommand*\EndOfBibitem{} \mciteSetBstSublistMode{f}
\mciteSetBstMaxWidthForm{subitem}{(\alph{mcitesubitemcount})}
\mciteSetBstSublistLabelBeginEnd
  {\mcitemaxwidthsubitemform\space}
  {\relax}
  {\relax}

\bibitem[Novoselov \latin{et~al.}(2004)Novoselov, Geim, Morozov, Jiang, Zhang,
  Dubonos, Grigorieva, and Firsov]{NovoselovSci04}
Novoselov,~K.~S.; Geim,~A.~K.; Morozov,~S.~V.; Jiang,~D.; Zhang,~Y.;
  Dubonos,~S.~V.; Grigorieva,~I.~V.; Firsov,~A.~A. \emph{Science}
  \textbf{2004}, \emph{306}, 666--669\relax
\mciteBstWouldAddEndPuncttrue
\mciteSetBstMidEndSepPunct{\mcitedefaultmidpunct}
{\mcitedefaultendpunct}{\mcitedefaultseppunct}\relax \EndOfBibitem
\bibitem[Fiori \latin{et~al.}(2014)Fiori, Bonaccorso, Iannaccone, T.Palacios,
  Neumaier, Seabaugh, Banerjee, and Colombo]{FioriNatNano14}
Fiori,~G.; Bonaccorso,~F.; Iannaccone,~G.; T.Palacios,;
Neumaier,~D.;
  Seabaugh,~A.; Banerjee,~S.~K.; Colombo,~L. \emph{Nat. Nanotech.}
  \textbf{2014}, \emph{10}, 768--779\relax
\mciteBstWouldAddEndPuncttrue
\mciteSetBstMidEndSepPunct{\mcitedefaultmidpunct}
{\mcitedefaultendpunct}{\mcitedefaultseppunct}\relax \EndOfBibitem
\bibitem[Koppens \latin{et~al.}(2014)Koppens, Mueller, Avouris, Ferrari,
  Vitiello, and Polini]{KoppensNatNano14}
Koppens,~F. H.~L.; Mueller,~T.; Avouris,~P.; Ferrari,~A.~C.;
Vitiello,~M.~S.;
  Polini,~M. \emph{Nat. Nanotech.} \textbf{2014}, \emph{10}, 780--793\relax
\mciteBstWouldAddEndPuncttrue
\mciteSetBstMidEndSepPunct{\mcitedefaultmidpunct}
{\mcitedefaultendpunct}{\mcitedefaultseppunct}\relax \EndOfBibitem
\bibitem[Mak \latin{et~al.}(2010)Mak, Lee, Hone, Shan, and Heinz]{MakPRL10}
Mak,~K.~F.; Lee,~C.; Hone,~J.; Shan,~J.; Heinz,~T.~F. \emph{Phys.
Rev. Lett.}
  \textbf{2010}, \emph{105}, 136805\relax
\mciteBstWouldAddEndPuncttrue
\mciteSetBstMidEndSepPunct{\mcitedefaultmidpunct}
{\mcitedefaultendpunct}{\mcitedefaultseppunct}\relax \EndOfBibitem
\bibitem[Deveaud \latin{et~al.}(1991)Deveaud, Cl\'{e}rot, Roy, Satzke, Sermage,
  and Katzer]{DeveaudPRL91}
Deveaud,~B.; Cl\'{e}rot,~F.; Roy,~N.; Satzke,~K.; Sermage,~B.;
Katzer,~D.~S.
  \emph{Phys. Rev. Lett.} \textbf{1991}, \emph{67}, 2355\relax
\mciteBstWouldAddEndPuncttrue
\mciteSetBstMidEndSepPunct{\mcitedefaultmidpunct}
{\mcitedefaultendpunct}{\mcitedefaultseppunct}\relax \EndOfBibitem
\bibitem[Ye \latin{et~al.}(2014)Ye, Cao, O'Brien, Zhu, Yin, Wang, Louie and Zhang]{YeNature14} Ye,~Z.; Cao,~T.; O'Brien,~K.; Zhu,~H.; Yin,~X.; Wang,~Y.;
Louie,~S.G.; Zhang,~X.; \emph{Nature} \textbf{2014}, \emph{513},
214--218\relax \mciteBstWouldAddEndPuncttrue
\mciteSetBstMidEndSepPunct{\mcitedefaultmidpunct}
{\mcitedefaultendpunct}{\mcitedefaultseppunct}\relax \EndOfBibitem
\bibitem[Chernikov \latin{et~al.}(2014)Chernikov, Berkelbach, Hill, Rigosi, Li, Aslan, Reichman, Hybertsen and
Heinz]{ChernikovPRL14} Chernikov,~A.; Berkelbach,~T.C.; Hill,~H.M.;
Rigosi,~A.; Li,~Y.; Aslan,~O.B.; Reichman,~D.R.; Hybertsen,~M.S.;
Heinz,~T.F. \emph{Phys. Rev. Lett.} \textbf{2014}, \emph{113},
076802\relax \mciteBstWouldAddEndPuncttrue
\mciteSetBstMidEndSepPunct{\mcitedefaultmidpunct}
{\mcitedefaultendpunct}{\mcitedefaultseppunct}\relax \EndOfBibitem
\bibitem[Wang \latin{et~al.}(2015)Wang, Marie, Gerber, Amand, Lagarde, Bouet, Vidal, Balocchi and
Urbaszek]{WangPRL15} Wang,~G.; Marie,~X.; Gerber,~I.; Amand,~T.;
Lagarde,~D.; Bouet,~L.; Vidal,~M.; Balocchi,~A.; Urbaszek,~B.
\emph{Phys. Rev. Lett.} \textbf{2015}, \emph{114}, 097403\relax
\mciteBstWouldAddEndPuncttrue
\mciteSetBstMidEndSepPunct{\mcitedefaultmidpunct}
{\mcitedefaultendpunct}{\mcitedefaultseppunct}\relax \EndOfBibitem
\bibitem[Olsen \latin{et~al.}(2016)Olsen, Latini, Rasmussen, and
  Thygesen]{OlsenPRL16}
Olsen,~T.; Latini,~S.; Rasmussen,~F.; Thygesen,~K.~S. \emph{Phys.
Rev. Lett.}  \textbf{2016}, \emph{116}, 056401\relax
\mciteBstWouldAddEndPuncttrue
\mciteSetBstMidEndSepPunct{\mcitedefaultmidpunct}
{\mcitedefaultendpunct}{\mcitedefaultseppunct}\relax \EndOfBibitem
\bibitem[Leavitt and Little(1990)Leavitt, and Little]{LeavittPRB90}
Leavitt,~R.~P.; Little,~J.~W. \emph{Phys. Rev. B} \textbf{1990},
\emph{42},
  11774\relax
\mciteBstWouldAddEndPuncttrue
\mciteSetBstMidEndSepPunct{\mcitedefaultmidpunct}
{\mcitedefaultendpunct}{\mcitedefaultseppunct}\relax \EndOfBibitem
\bibitem[Andreani(1995)]{AndreaniBook94}
Andreani,~L.~C. In \emph{Confined Electrons and Photons: New Physics
and
  Applications}; Burstein,~E., Weisbuch,~C., Eds.; Nato Science Series: B
  Physics; Plenum Press: New York, 1995; Vol. 340; pp 57--112\relax
\mciteBstWouldAddEndPuncttrue
\mciteSetBstMidEndSepPunct{\mcitedefaultmidpunct}
{\mcitedefaultendpunct}{\mcitedefaultseppunct}\relax \EndOfBibitem
\bibitem[Arora \latin{et~al.}(2015)Arora, Koperski, Nogajewski, Marcus,
  Faugeras, and Potemski]{AroraNanoscale15}
Arora,~A.; Koperski,~M.; Nogajewski,~K.; Marcus,~J.; Faugeras,~C.;
Potemski,~M.
  \emph{Nanoscale} \textbf{2015}, \emph{7}, 10421--10429\relax
\mciteBstWouldAddEndPuncttrue
\mciteSetBstMidEndSepPunct{\mcitedefaultmidpunct}
{\mcitedefaultendpunct}{\mcitedefaultseppunct}\relax \EndOfBibitem
\bibitem[Palummo \latin{et~al.}(2015)Palummo, Bernardi, and
  Grossman]{PalummoNL15}
Palummo,~M.; Bernardi,~M.; Grossman,~J.~C. \emph{Nano Lett.}
\textbf{2015},
  \emph{15}, 2794\relax
\mciteBstWouldAddEndPuncttrue
\mciteSetBstMidEndSepPunct{\mcitedefaultmidpunct}
{\mcitedefaultendpunct}{\mcitedefaultseppunct}\relax \EndOfBibitem
\bibitem[Poellmann \latin{et~al.}(2015)Poellmann, Steinleitner, Leierseder,
  Nagler, Plechinger, Porer, Bratschitsch, Sch\"{u}ller, Korn, and
  Huber]{PoellmannNatMat15}
Poellmann,~C.; Steinleitner,~P.; Leierseder,~U.; Nagler,~P.;
Plechinger,~G.;
  Porer,~M.; Bratschitsch,~R.; Sch\"{u}ller,~C.; Korn,~T.; Huber,~R. \emph{Nat.
  Mater.} \textbf{2015}, \emph{14}, 889--893\relax
\mciteBstWouldAddEndPuncttrue
\mciteSetBstMidEndSepPunct{\mcitedefaultmidpunct}
{\mcitedefaultendpunct}{\mcitedefaultseppunct}\relax \EndOfBibitem
\bibitem[Moody \latin{et~al.}(2015)Moody, Dass, Hao, Chen, Li, Singh, Tran,
  Clark, Xu, Bergha\"{u}ser, Malic, Knorr, and Li]{MoodyNatComm15}
Moody,~G.; Dass,~C.~K.; Hao,~K.; Chen,~C.-H.; Li,~L.-J.; Singh,~A.;
Tran,~K.;
  Clark,~G.; Xu,~X.; Bergha\"{u}ser,~G.; Malic,~E.; Knorr,~A.; Li,~X.
  \emph{Nat. Comm.} \textbf{2015}, \emph{6}, 8315\relax
\mciteBstWouldAddEndPuncttrue
\mciteSetBstMidEndSepPunct{\mcitedefaultmidpunct}
{\mcitedefaultendpunct}{\mcitedefaultseppunct}\relax \EndOfBibitem
\bibitem[Liu \latin{et~al.}(2014)Liu, Galfsky, Sun, Xia, chen Lin, Lee,
  K\'{e}na-Cohen, and Menon]{LiuNatPhot14}
Liu,~X.; Galfsky,~T.; Sun,~Z.; Xia,~F.; chen Lin,~E.; Lee,~Y.-H.;
  K\'{e}na-Cohen,~S.; Menon,~V.~M. \emph{Nat. Phot.} \textbf{2014}, \emph{9},
  30--34\relax
\mciteBstWouldAddEndPuncttrue
\mciteSetBstMidEndSepPunct{\mcitedefaultmidpunct}
{\mcitedefaultendpunct}{\mcitedefaultseppunct}\relax \EndOfBibitem
\bibitem[Malard \latin{et~al.}(2013)Malard, Alencar, Barboza, Mak, and
  de~Paula]{MalardPRB13}
Malard,~L.~M.; Alencar,~T.~V.; Barboza,~A. P.~M.; Mak,~K.~F.;
de~Paula,~A.~M.
  \emph{Phs. Rev. B} \textbf{2013}, \emph{87}, 201401(R)\relax
\mciteBstWouldAddEndPuncttrue
\mciteSetBstMidEndSepPunct{\mcitedefaultmidpunct}
{\mcitedefaultendpunct}{\mcitedefaultseppunct}\relax \EndOfBibitem
\bibitem[Lagarde \latin{et~al.}(2014)Lagarde, Bouet, Marie, Zhu, Liu, Amand,
  Tan, and Urbaszek]{LagardePRL14}
Lagarde,~D.; Bouet,~L.; Marie,~X.; Zhu,~C.~R.; Liu,~B.~L.;
Amand,~T.;
  Tan,~P.~H.; Urbaszek,~B. \emph{Phys. Rev. Lett.} \textbf{2014}, \emph{112},
  047401\relax
\mciteBstWouldAddEndPuncttrue
\mciteSetBstMidEndSepPunct{\mcitedefaultmidpunct}
{\mcitedefaultendpunct}{\mcitedefaultseppunct}\relax \EndOfBibitem
\bibitem[Hao \latin{et~al.}(2015)Hao, Moody, Wu, Dass, Xu, Chen, Li, Li,
  MacDonald, and Li]{Hao2015}
Hao,~K.; Moody,~G.; Wu,~F.; Dass,~C.~K.; Xu,~L.; Chen,~C.-H.;
Li,~M.-Y.;
  Li,~L.-J.; MacDonald,~A.~H.; Li,~X. \emph{Nat. Phys.} \emph{doi:10.1038/nphys3674} \textbf{2016},
  \relax
\mciteBstWouldAddEndPunctfalse
\mciteSetBstMidEndSepPunct{\mcitedefaultmidpunct}
{}{\mcitedefaultseppunct}\relax \EndOfBibitem
\bibitem[Fras \latin{et~al.}(2015)Fras, Mermillod, Nogues, Hoarau, Schneider,
  Kamp, H\"{o}fing, Langbein, and Kasprzak]{FrasNatPhot16}
Fras,~F.; Mermillod,~Q.; Nogues,~G.; Hoarau,~C.; Schneider,~C.;
Kamp,~M.;
  H\"{o}fing,~S.; Langbein,~W.; Kasprzak,~J. \emph{Nat. Phot.} \textbf{2016},
  \emph{10}, 155\relax
\mciteBstWouldAddEndPuncttrue
\mciteSetBstMidEndSepPunct{\mcitedefaultmidpunct}
{\mcitedefaultendpunct}{\mcitedefaultseppunct}\relax \EndOfBibitem
\bibitem[Langbein and Patton(2005)Langbein, and Patton]{LangbeinPRL05}
Langbein,~W.; Patton,~B. \emph{Phys. Rev. Lett.} \textbf{2005},
\emph{95},
  017403\relax
\mciteBstWouldAddEndPuncttrue
\mciteSetBstMidEndSepPunct{\mcitedefaultmidpunct}
{\mcitedefaultendpunct}{\mcitedefaultseppunct}\relax \EndOfBibitem
\bibitem[Wang \latin{et~al.}(2015)Wang, Palleau, Amand, Tongay, Marie, and
  Urbaszek]{WangAPL15}
Wang,~G.; Palleau,~E.; Amand,~T.; Tongay,~S.; Marie,~X.;
Urbaszek,~B.
  \emph{Appl. Phys. Lett.} \textbf{2015}, \emph{106}, 112101\relax
\mciteBstWouldAddEndPuncttrue
\mciteSetBstMidEndSepPunct{\mcitedefaultmidpunct}
{\mcitedefaultendpunct}{\mcitedefaultseppunct}\relax \EndOfBibitem
\bibitem[Jones \latin{et~al.}(2016)Jones, Yu, Ghimire, Wu, Aivazian, Ross,
  Zhao, Yan, Mandrus, Xiao, Yao, , and Xu]{JonesNatNano13}
Jones,~A.~M.; Yu,~H.; Ghimire,~N.~J.; Wu,~S.; Aivazian,~G.;
Ross,~J.~S.;
  Zhao,~B.; Yan,~J.; Mandrus,~D.~G.; Xiao,~D.; Yao,~W.; ; Xu,~X. \emph{Nat.
  Nanotech.} \textbf{2016}, \emph{8}, 634\relax
\mciteBstWouldAddEndPuncttrue
\mciteSetBstMidEndSepPunct{\mcitedefaultmidpunct}
{\mcitedefaultendpunct}{\mcitedefaultseppunct}\relax \EndOfBibitem
\bibitem[Langbein \latin{et~al.}(2002)Langbein, Runge, Savona, and
  Zimmermann]{LangbeinPRL02a}
Langbein,~W.; Runge,~E.; Savona,~V.; Zimmermann,~R. \emph{Phys. Rev.
Lett.}
  \textbf{2002}, \emph{89}, 157401\relax
\mciteBstWouldAddEndPuncttrue
\mciteSetBstMidEndSepPunct{\mcitedefaultmidpunct}
{\mcitedefaultendpunct}{\mcitedefaultseppunct}\relax \EndOfBibitem
\bibitem[Kasprzak \latin{et~al.}(2011)Kasprzak, Patton, Savona, and
  Langbein]{KasprzakNPho11}
Kasprzak,~J.; Patton,~B.; Savona,~V.; Langbein,~W. \emph{Nat. Phot.}
  \textbf{2011}, \emph{5}, 57-63\relax
\mciteBstWouldAddEndPuncttrue
\mciteSetBstMidEndSepPunct{\mcitedefaultmidpunct}
{\mcitedefaultendpunct}{\mcitedefaultseppunct}\relax \EndOfBibitem
\bibitem[Langbein(2010)]{LangbeinRNC10}
Langbein,~W. \emph{Rivista del nuovo cimento} \textbf{2010},
\emph{33},
  255--312\relax
\mciteBstWouldAddEndPuncttrue
\mciteSetBstMidEndSepPunct{\mcitedefaultmidpunct}
{\mcitedefaultendpunct}{\mcitedefaultseppunct}\relax \EndOfBibitem
\bibitem[Naeem \latin{et~al.}(2015)Naeem, Masia, Christodoulou, Moreels, Borri,
  and Langbein]{NaeemPRB15}
Naeem,~A.; Masia,~F.; Christodoulou,~S.; Moreels,~I.; Borri,~P.;
Langbein,~W.
  \emph{Phys. Rev. B} \textbf{2015}, \emph{91}, 121302(R)\relax
\mciteBstWouldAddEndPuncttrue
\mciteSetBstMidEndSepPunct{\mcitedefaultmidpunct}
{\mcitedefaultendpunct}{\mcitedefaultseppunct}\relax \EndOfBibitem
\bibitem[Singh \latin{et~al.}(2014)Singh, Moody, Wu, Wu, Ghimire, Yan, Mandrus,
  Xu, and Li]{SinghPRL14}
Singh,~A.; Moody,~G.; Wu,~S.; Wu,~Y.; Ghimire,~N.~J.; Yan,~J.;
Mandrus,~D.~G.;
  Xu,~X.; Li,~X. \emph{Phys. Rev. Lett.} \textbf{2014}, \emph{112},
  216804\relax
\mciteBstWouldAddEndPuncttrue
\mciteSetBstMidEndSepPunct{\mcitedefaultmidpunct}
{\mcitedefaultendpunct}{\mcitedefaultseppunct}\relax \EndOfBibitem
\bibitem[Porras \latin{et~al.}(2002)Porras, Ciuti, Baumberg, and
  Tejedor]{PorrasPRB02}
Porras,~D.; Ciuti,~C.; Baumberg,~J.~J.; Tejedor,~C. \emph{Phys. Rev.
B}
  \textbf{2002}, \emph{66}, 085304\relax
\mciteBstWouldAddEndPuncttrue
\mciteSetBstMidEndSepPunct{\mcitedefaultmidpunct}
{\mcitedefaultendpunct}{\mcitedefaultseppunct}\relax \EndOfBibitem
\bibitem[Masia \latin{et~al.}(2011)]{MasiaPRB11}
Masia,~F.; Langbein,~W.; Moreels,~I.; Hens,~Z.; and Borri,~P.
\emph{Phys. Rev. B} \textbf{2011},
  \emph{83}, 201305(R)\relax
\mciteBstWouldAddEndPuncttrue
\mciteSetBstMidEndSepPunct{\mcitedefaultmidpunct}
{\mcitedefaultendpunct}{\mcitedefaultseppunct}\relax \EndOfBibitem
\bibitem[An \latin{et~al.}(2011)]{AnNL07}
An,~J.~M.; Franceschetti,~A.; and Zunger,~A. \emph{Nano. Lett.}
\textbf{2007},
  \emph{7}, 2129\relax
\mciteBstWouldAddEndPuncttrue
\mciteSetBstMidEndSepPunct{\mcitedefaultmidpunct}
{\mcitedefaultendpunct}{\mcitedefaultseppunct}\relax \EndOfBibitem
\bibitem[Qiu \latin{et~al.}(2015)Qiu, Cao, and Louie]{QiuPRL15}
Qiu,~D.~Y.; Cao,~T.; Louie,~S.~G. \emph{Phys. Rev. Lett.}
\textbf{2015},
  \emph{115}, 176801\relax
\mciteBstWouldAddEndPuncttrue
\mciteSetBstMidEndSepPunct{\mcitedefaultmidpunct}
{\mcitedefaultendpunct}{\mcitedefaultseppunct}\relax \EndOfBibitem
\bibitem[Xiao \latin{et~al.}(2012)Xiao, Liu, Feng, Xu, and Yao]{XiaoPRL2012}
Xiao,~D.; Liu,~G.-B.; Feng,~W.; Xu,~X.; Yao,~W. \emph{Phys. Rev.
Lett.}
  \textbf{2012}, \emph{108}, 196802\relax
\mciteBstWouldAddEndPuncttrue
\mciteSetBstMidEndSepPunct{\mcitedefaultmidpunct}
{\mcitedefaultendpunct}{\mcitedefaultseppunct}\relax \EndOfBibitem
\bibitem[Jones \latin{et~al.}(2013)Jones, Yu, Ghimire, SanfengWu, Aivazian,
  Ross, Zhao, Yan, Mandrus, Xiao, Yao, and Xu]{JonesNatNano2013}
Jones,~A.~M.; Yu,~H.; Ghimire,~N.~J.; SanfengWu,; Aivazian,~G.;
Ross,~J.~S.;
  Zhao,~B.; Yan,~J.; Mandrus,~D.~G.; Xiao,~D.; Yao,~W.; Xu,~X. \emph{Nat.
  Nanotech.} \textbf{2013}, \emph{8}, 634--638\relax
\mciteBstWouldAddEndPuncttrue
\mciteSetBstMidEndSepPunct{\mcitedefaultmidpunct}
{\mcitedefaultendpunct}{\mcitedefaultseppunct}\relax \EndOfBibitem
\bibitem[Xu \latin{et~al.}(2014)Xu, Yao, Xiao, and Heinz]{XuNatPhys14}
Xu,~X.; Yao,~W.; Xiao,~D.; Heinz,~T.~F. \emph{Nat. Phys.}
\textbf{2014},
  \emph{10}, 343--350\relax
\mciteBstWouldAddEndPuncttrue
\mciteSetBstMidEndSepPunct{\mcitedefaultmidpunct}
{\mcitedefaultendpunct}{\mcitedefaultseppunct}\relax \EndOfBibitem
\bibitem[Yu and Wu(2014)Yu, and Wu]{YuPRB14}
Yu,~T.; Wu,~M.~W. \emph{Phys. Rev. B} \textbf{2014}, \emph{89},
205303\relax \mciteBstWouldAddEndPuncttrue
\mciteSetBstMidEndSepPunct{\mcitedefaultmidpunct}
{\mcitedefaultendpunct}{\mcitedefaultseppunct}\relax \EndOfBibitem
\bibitem[Glazov \latin{et~al.}(2014)Glazov, Amand, Marie, Lagarde, Bouet, and
  Urbaszek]{GlazovPRB14}
Glazov,~M.~M.; Amand,~T.; Marie,~X.; Lagarde,~D.; Bouet,~L.;
Urbaszek,~B.
  \emph{Phys. Rev. B} \textbf{2014}, \emph{89}, 201302(R)\relax
\mciteBstWouldAddEndPuncttrue
\mciteSetBstMidEndSepPunct{\mcitedefaultmidpunct}
{\mcitedefaultendpunct}{\mcitedefaultseppunct}\relax \EndOfBibitem
\bibitem[Smole\'{n}ski \latin{et~al.}(2016)Smole\'{n}ski, Goryca, Koperski,
  Faugeras, Kazimierczuk, Nogajewski, Kossacki, and Potemski]{Smolenski16}
Smole\'{n}ski,~T.; Goryca,~M.; Koperski,~M.; Faugeras,~C.;
Kazimierczuk,~T.;
  Nogajewski,~K.; Kossacki,~P.; Potemski,~M. \emph{Phys. Rev. X}
  \textbf{2016}, \emph{6}, 021024\relax
\mciteBstWouldAddEndPunctfalse
\mciteSetBstMidEndSepPunct{\mcitedefaultmidpunct}
{}{\mcitedefaultseppunct}\relax \EndOfBibitem
\bibitem[Zhang \latin{et~al.}(2015)Zhang, You, Zhao, and Heinz]{ZhangPRL15}
Zhang,~X.-X.; You,~Y.; Zhao,~S. Y.~F.; Heinz,~T.~F. \emph{Phys. Rev.
Lett.}
  \textbf{2015}, \emph{115}, 257403\relax
\mciteBstWouldAddEndPuncttrue
\mciteSetBstMidEndSepPunct{\mcitedefaultmidpunct}
{\mcitedefaultendpunct}{\mcitedefaultseppunct}\relax \EndOfBibitem
\bibitem[Wang \latin{et~al.}(2015)Wang, Robert, Suslu, Chen, Yang, Alamdari,
  Gerber, Amand, Marie, Tongay, and Urbaszek]{WangNatComm2015}
Wang,~G.; Robert,~C.; Suslu,~A.; Chen,~B.; Yang,~S.; Alamdari,~S.;
  Gerber,~I.~C.; Amand,~T.; Marie,~X.; Tongay,~S.; Urbaszek,~B. \emph{Nat.
  Comm.} \textbf{2015}, \emph{6}, 10110\relax
\mciteBstWouldAddEndPuncttrue
\mciteSetBstMidEndSepPunct{\mcitedefaultmidpunct}
{\mcitedefaultendpunct}{\mcitedefaultseppunct}\relax \EndOfBibitem
\bibitem[Horzum \latin{et~al.}(2013)Horzum, Sahin, Cahangirov, Cudazzo, Rubio, Serin, and Peeters]{HorzumPRB13}
Horzum,~S.; Sahin,~H.; Cahangirov,~S.; Cudazzo,~P.; Rubio,~A.;
Serin,~T.; Peeters.~F.M.; \emph{Phys. Rev. B.} \textbf{2013},
\emph{87}, 125415\relax \mciteBstWouldAddEndPuncttrue
\mciteSetBstMidEndSepPunct{\mcitedefaultmidpunct}
{\mcitedefaultendpunct}{\mcitedefaultseppunct}\relax \EndOfBibitem
\bibitem[Wang \latin{et~al.}(2015)Wang, Robert, Suslu, Chen, Yang, Alamdari,
  Gerber, Amand, Marie, Tongay, and Urbaszek]{WangNatComm2015}
Wang,~G.; Robert,~C.; Suslu,~A.; Chen,~B.; Yang,~S.; Alamdari,~S.;
  Gerber,~I.~C.; Amand,~T.; Marie,~X.; Tongay,~S.; Urbaszek,~B. \emph{Nat.
  Comm.} \textbf{2015}, \emph{6}, 10110\relax
\mciteBstWouldAddEndPuncttrue
\mciteSetBstMidEndSepPunct{\mcitedefaultmidpunct}
{\mcitedefaultendpunct}{\mcitedefaultseppunct}\relax \EndOfBibitem
\bibitem[Echeverry \latin{et~al.}(2016)Echeverry, Urbaszek, Amand, Marie, and
  Gerber]{Echeverry2016}
Echeverry,~J.~P.; Urbaszek,~B.; Amand,~T.; Marie,~X.; Gerber,~I.~C.
  \emph{arXiv:1601.07351} \textbf{2016}, \relax
\mciteBstWouldAddEndPunctfalse
\mciteSetBstMidEndSepPunct{\mcitedefaultmidpunct}
{}{\mcitedefaultseppunct}\relax \EndOfBibitem
\bibitem[Withers \latin{et~al.}(2015)Withers, Pozo-Zamudio, Schwarz,
  Dufferwiel, Walker, Godde, Rooney, Gholinia, Woods, Blake, Haigh, Watanabe,
  Taniguchi, Aleiner, Geim, Fal'ko, Tartakovskii, and
  Novoselov]{WithersNanoLett15}
Withers,~F. \latin{et~al.}  \emph{Nano Lett.} \textbf{2015},
\emph{15},
  8223--8228\relax
\mciteBstWouldAddEndPuncttrue
\mciteSetBstMidEndSepPunct{\mcitedefaultmidpunct}
{\mcitedefaultendpunct}{\mcitedefaultseppunct}\relax \EndOfBibitem
\bibitem[Dery and Song(2015) Hanan Dery and Yang Song]{DeryPRB15}
Dery,~H.; and Song,~Y. \emph{Phys. Rev. B} \textbf{2015}, \emph{92},
125431\relax \mciteBstWouldAddEndPuncttrue
\mciteSetBstMidEndSepPunct{\mcitedefaultmidpunct}
{\mcitedefaultendpunct}{\mcitedefaultseppunct}\relax \EndOfBibitem
\bibitem[Rivera \latin{et~al.}(2015)Rivera, Schaibley, Jones, Ross, Wu,
  Aivazian, Klement, Seyler, Clark, Ghimire, Yan, Mandrus, Yao, and
  Xu]{RiveraNatComm15}
Rivera,~P.; Schaibley,~J.~R.; Jones,~A.~M.; Ross,~J.~S.; Wu,~S.;
Aivazian,~G.;
  Klement,~P.; Seyler,~K.; Clark,~G.; Ghimire,~N.~J.; Yan,~J.; Mandrus,~D.~G.;
  Yao,~W.; Xu,~X. \emph{Nat. Comm.} \textbf{2015}, \emph{6}, 6242\relax
\mciteBstWouldAddEndPuncttrue
\mciteSetBstMidEndSepPunct{\mcitedefaultmidpunct}
{\mcitedefaultendpunct}{\mcitedefaultseppunct}\relax \EndOfBibitem
\bibitem[Rivera \latin{et~al.}(2016)Rivera, Seyler, Yu, Schaibley, Yan,
  Mandrus, Yao, and Xu]{Rivera16}
Rivera,~P.; Seyler,~K.~L.; Yu,~H.; Schaibley,~J.~R.; Yan,~J.;
Mandrus,~D.~G.;
  Yao,~W.; Xu,~X. \emph{arXiv:1601.02641} \textbf{2016}, \relax
\mciteBstWouldAddEndPunctfalse
\mciteSetBstMidEndSepPunct{\mcitedefaultmidpunct}
{}{\mcitedefaultseppunct}\relax \EndOfBibitem
\bibitem[Koperski \latin{et~al.}(2015)Koperski, Nogajewski, Arora, Cherkez,
  Mallet, Veuillen, Marcus, Kossacki, and Potemski]{KoperskiNN2015}
Koperski,~M.; Nogajewski,~K.; Arora,~A.; Cherkez,~V.; Mallet,~P.;
  Veuillen,~J.-Y.; Marcus,~J.; Kossacki,~P.; Potemski,~M. \emph{Nat. Nanotech.}
  \textbf{2015}, \emph{10}, 503--506\relax
\mciteBstWouldAddEndPuncttrue
\mciteSetBstMidEndSepPunct{\mcitedefaultmidpunct}
{\mcitedefaultendpunct}{\mcitedefaultseppunct}\relax \EndOfBibitem
\bibitem[Schaibley \latin{et~al.}(2015)Schaibley, Karin, Yu, Ross, Rivera,
  Jones, Scott, Yan, Mandrus, Yao, Fu, and Xu]{SchaibleyPRL15}
Schaibley,~J.~R.; Karin,~T.; Yu,~H.; Ross,~J.~S.; Rivera,~P.;
Jones,~A.~M.;
  Scott,~M.~E.; Yan,~J.; Mandrus,~D.~G.; Yao,~W.; Fu,~K.-M.; Xu,~X. \emph{Phys.
  Rev. Lett.} \textbf{2015}, \emph{114}\relax
\mciteBstWouldAddEndPuncttrue
\mciteSetBstMidEndSepPunct{\mcitedefaultmidpunct}
{\mcitedefaultendpunct}{\mcitedefaultseppunct}\relax \EndOfBibitem
\end{mcitethebibliography}
\end{document}